# Super Scatterers Based on Artificial Localized Magnon Resonances

Dmitry Filonov[1,‡], Andrey Shmidt[1], Amir Boag[1] and Pavel Ginzburg[1,2]

[1]School of Electrical Engineering, Tel Aviv University, Tel Aviv, 69978, Israel

[2]Light-Matter Interaction Centre, Tel Aviv University, Tel Aviv, 69978, Israel

**Abstract**: The interaction between electromagnetic waves and objects is strongly affected by the shape and material composition of the latter. Artificially created materials, formed by a subwavelength structuring of their unit cells, namely metamaterials, can exhibit peculiar responses to electromagnetic radiation and provide additional powerful degrees of freedom to the scatterer design. In particular, negative material susceptibilities give rise to strong resonant interactions with deeply subwavelength particles. While the negative electrical permittivity of natural noble metals manifests itself in localized plasmon resonant oscillations, negative magnetic permeability is virtually non-existent in nature. Here the concept of artificial magnon resonance in subwavelength objects with effective negative permeability, designed based on the metamaterial approach, is demonstrated. Strong localized oscillations of the magnetic fields within an array of split ring resonators, forming a sphere, hybridize in a collective mode of the structure. As a result, extremely high scattering cross section, exceeding that of a steel sphere with the same radius by four orders of magnitude, was demonstrated. Furthermore, the scattering cross section of subwavelength metamaterial-based sphere was shown to be comparable to the low frequency (MHz) radar signature of a big military aircraft. Super scatterers, based on tunable resonances within artificially created materials, can find use in a broad range of electromagnetic applications, including wireless communications, radars, RFID, internet of things hardware and many others.

[‡][dimfilonov@gmail.com](mailto:dimfilonov@gmail.com)

**Introduction**

Interactions of electromagnetic waves with material structures give rise to many fundamental phenomena, occurring over the whole spectral range (e.g. optical [1] and low GHz-MHz [2]) and give rise to a vast variety of nowadays applications. Scattering cross sections of objects are solely defined by their geometries and material compositions. A span of optical materials is rather broad, which allows considering the material degree of freedom as a design parameter. For example, noble metals, such as silver and gold, have moderate negative permittivity at visible and infrared spectral ranges. This property enables observing and controlling resonant scattering phenomena from deeply subwavelength geometries and it is referred as localized plasmon resonance (LPR), e.g. [3],[4]. However, metals exhibit extreme negative epsilon behavior and become primarily conductive at low-frequency ranges, virtually eliminating the ability to support LPR phenomenon. Consequently, achieving resonant behavior of structures at MHz-GHz frequency ranges requires the utilization of retardation effects, which imposes limitations on linear dimensions of scatterers [2]. Reducing the physical sizes of antennas, employed in wireless communication links, is a long-standing task which requires exploring tradeoffs between many parameters, pre-defining the performance. However, material properties, in the majority of the cases, are not considered as valuable degrees of freedom, which can bring qualitatively new concepts into the field. As an exception, it is worth mentioning high epsilon ceramic composites [5], which enable shrinking effective wavelengths by orders of magnitude. Those materials, however, require complex fabrication routines and hardly sustain post processing due to their brittle nature. One of the very promising concepts for tailoring material properties, which was found to be successful in applications in the MHz-THz frequency ranges, is to develop arrays and surfaces with subwavelength structuring. Properties of those artificially created composites can be homogenized and encapsulated within effective permittivity and permeability parameters – this is the concept of metamaterials [6],[7],[8],[9],[10]. Radio frequency (RF) metamaterials were suggested and demonstrated for invisibility cloaking applications [11], scattering suppression [12], negative refraction [13], imaging with super-resolution [14], metamaterial-based cavities [15], compact cavity resonators [16], emulation tools for complex optical phenomena (e.g. [17]), antennas with improved characteristics (e.g. [18],[19],[20]) and many others. Negative permittivity and permeability (in separate) composites are especially relevant to this Communication Letter. For example, wire medium was proposed to serve as an artificial plasmonic material ($\varepsilon<0$) at RF [21], while strong spatial dispersion should be taken into account in order to extract effective material parameters [22]. Other type (resonant) of artificial RF plasmonic materials were demonstrated in application to the invisibility cloaking [23]. While realizations of strong collective electric responses require employment of dipolar-like resonators within an array, magnetic phenomena are governed by microscopic (on the subwavelength scale in the context of metamaterials) circular currents. Split ring resonators (SRRs) are frequently employed for achieving artificial magnetic responses, e.g. [24],[25]. Limitations of the homogenization theory in application to the analysis of finite size SRR-based geometries were recently analyzed and relatively small deviations were shown to emerge owing to modifications in boundary conditions [26]. If the collective response of near-field coupled SRRs (or other types of resonators) is strong enough, the homogenized effective permeability can become negative. As a result, small subwavelength structures can exhibit strong localized resonances, driven by magnetic fields, in a similarity with LPR. This effect will be investigated hereafter and taken towards applications, where high Radar Cross Sections (RCSs) are required.

This Communication Letter is organized as follows: the effect of an artificial localized magnon resonance will be introduced and homogenization procedures towards achieving negative effective permeability in SRRs arrays will be introduced at first. A design of a deeply subwavelength metamaterial-based sphere, operating in the MHz frequency range, will follow. It will be shown that a small metamaterial-based object, as small as 10cm in diameter, can have a VHF radar signature, comparable with those of big aircrafts. The chart diagram in Fig. 1 summarizes the comparison between different RCSs, showing the strength of metamaterial-based resonators in scattering applications. The design, full numerical analysis and experimental study in the GHz spectral range will come before the 'Outlook and Conclusions'.

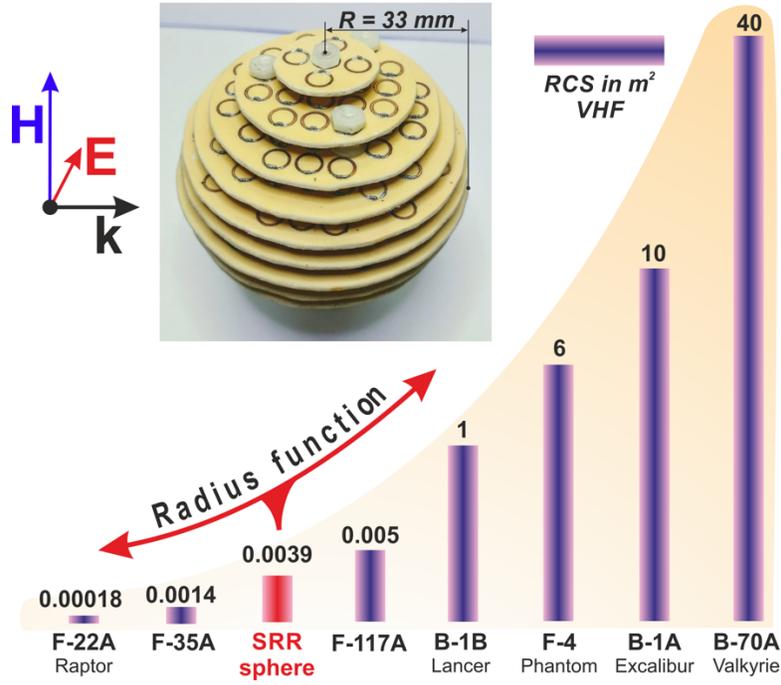

*Fig. 1. (Inset) Photograph of the metamaterial-based spherical scatterer (33 mm radius), comprising split ring resonator arrays, loaded with lumped capacitors. (Main figure) Bar chart, demonstrating VHF radar cross-section (RCS) of several aircrafts [27]. Numbers on top of the bars correspond to the values of RCS in m², Red bar corresponds to the meta-scatterer, whose RCS can be further controlled by changing its radius.*

**Negative permeability composites and artificial localized magnon resonances**

Localized plasmon resonance phenomenon utilizes collective oscillations of conduction electrons within noble metals and causes small deeply subwavelength particles to exhibit strong resonant responses. For example, 5 nm gold sphere resonates at 500 nm, demonstrating performance of an efficient electromagnetically small resonator, which is as small as λ/100 in size [3]. Relying on the duality of electric and magnetic fields, intrinsic to the Maxwell's equations, a similar phenomenon is expected to emerge in the case of negative permeability structures. In this case, the polarizability of a small magnetic sphere is given by:

$$\alpha_m = 4\pi r^3 \left(\frac{\mu_r - 1}{\mu_r + 2}\right), \tag{1}$$

where $\mu_r$ is the relative permeability of the material, while $r$ is the radius of the sphere. This expression can be modified to include the retardation effects, if the skin depth (for magnetic field penetration) becomes comparable with the radius. With the full analogy to the electric particles [28],[29], the modified polarizability is given by:

$$\alpha_m = 4\pi \left( \frac{1}{r^3} \cdot \frac{\mu_r + 2}{\mu_r - 1} - i\frac{2}{3}k^3 - \frac{k^2}{r} \right)^{-1}, \tag{2}$$

where $k$ is the free space wavenumber of the incident radiation. Eq. 1 explicitly shows that the resonant conditions are obtained when $Re(\mu_r) = -2$ and the quality factor of the resonance is limited by the material losses. Following the terminology of the localized plasmon resonance in the case of $Re(\varepsilon_r) = -2$, the magnetic phenomenon was coined here by the name of localized magnon resonances with the classification 'artificial' in order not to confuse it with the true solid state effect of collective excitation of the electrons' spins in a crystal lattice [30].

It is worth noting that the resonant condition of Eq. 1 strongly depends on the geometry and can be tuned by its adjustment, e.g. [4],[31]. Spherical geometry will be chosen hereafter only for the sake of the analytical simplicity of the analysis.

In contrast to negative permittivity, negative permeability materials are virtually non-existent in nature. Nevertheless, they can be created artificially by constructing arrays of ring resonators, which mimic a crystal of microscopic magnetic spins. Additional gap with or without a lumped capacitor element can be added within the ring in order to tune the resonance of the structure to a desired frequency [25]. The layout and parameters of the split ring resonator appears in the inset to Fig. 2(a). Identical SRRs were placed at the nodes of a 3D array (at the nodes of a simple cubic crystal with the period *a*) and effective electric and magnetic susceptibilities were retrieved. While an ideal array of rings can be homogenized analytically by applying coupled dipoles summations [32], a full wave numerical simulation with CST Microwave Studio was performed (frequency domain solver with $10^{-9}$ convergence parameter for very accurate modelling). It allows taking into account effects of asymmetry, impact of substrates and losses, and higher order electric and magnetic couplings, which are ignored in the beforehand mentioned analytical approach. Typical two-port waveguide system, with relevant dimensions to the specific frequency range was used. Complex-valued transmission and reflection coefficients (so called S-parameters) were calculated for three different orientations, corresponding to the main crystallographic axis of the metamaterial, and the permittivity and permeability tensors were extracted, following the formulation, reported at [33]. The following parameters of the structure were obtained after a set of optimizations: $r_{SRR}$ = 2.5 mm, $h$ = 0.5 mm, $a$ = 6 mm, $C$ = 220 pF. The substrate was taken to be *Isola IS680 AG338* ($\varepsilon$ = 3.38, $tg\delta$ = 0.0026) of 1.5 mm thickness, and the material of the SRR is copper (conductivity of 5.96×10$^7$ S/m). The resulting dispersion of effective magnetic medium appears in Fig. 2(a). Both real and imaginary parts resemble the classical Lorentzian shape close to the collective resonance. The region of interest, where the effective real part of the permeability approaches values of **-2** is highlighted with the yellow shadow. The corresponding frequency was chosen to fall within the region of VHF radar operation [34] for a demonstration of the concept only. It is worth emphasising, that the homogenised arrays of planar rings have strongly anisotropic permeability tensor – Fig. 2(a) demonstrates its component along the normal to the rings, while other two are virtually equal to unity. As the result, the composite behaves as a hyperbolic metamaterial [35] for the magnetic field. 'Hyperbolic' effects, however, do not play a role here and mentioned only owing to the special form of the effective tensor.

Scattering cross-section spectrum of the 5 cm sphere, made of the artificial magnetic material appears in Fig. 2(b). Two scenarios for the material permeability were considered here – isotropic $\mu_r$ and hyperbolic tensor, discussed beforehand. In deep subwavelength geometries, where the scattered field is known to be uniform inside the sphere, both of the cases demonstrate exactly the same behavior (polarization of the incident magnetic field is assumed to be along the direction of the nontrivial tensor component). The comparison between the meta-scatterer and a solid sphere, made of perfect electric conductor (PEC) appears in Fig. 2(b). Four orders of magnitude in the RCS enhancement owing to the artificial localized magnon resonance can be observed. Insets to panel b demonstrate the comparison of scattering diagrams on the linear scale. Note, that data for the PEC sphere is multiplied by the factor of 15 in order to make it visible. This tremendous enhancement is the core of the effect, schematically depicted in Fig. 1, where 33 mm sphere scatters as much as a big aircraft (33 mm is the actually fabricated device, while 5 cm was taken for numerical investigations and comparison). Note, that both of the objects are subwavelength in the case of VHF radar applications. It is also worth mentioning the advantage of the localized magnon resonance in scattering applications over high-quality resonators, made of high permittivity ceramic materials. For example, 5 cm radius sphere, made of BaTiO$_3$-SrTiO$_3$ with $\varepsilon = 533\ and\ tg\sigma = 0.018$ [36], have a dipolar resonance at 130 MHz. The RCS peak of this structure is 30 times smaller than the value, demonstrated by the metamaterial-based scatterer, underlining the strength of the proposed approach.

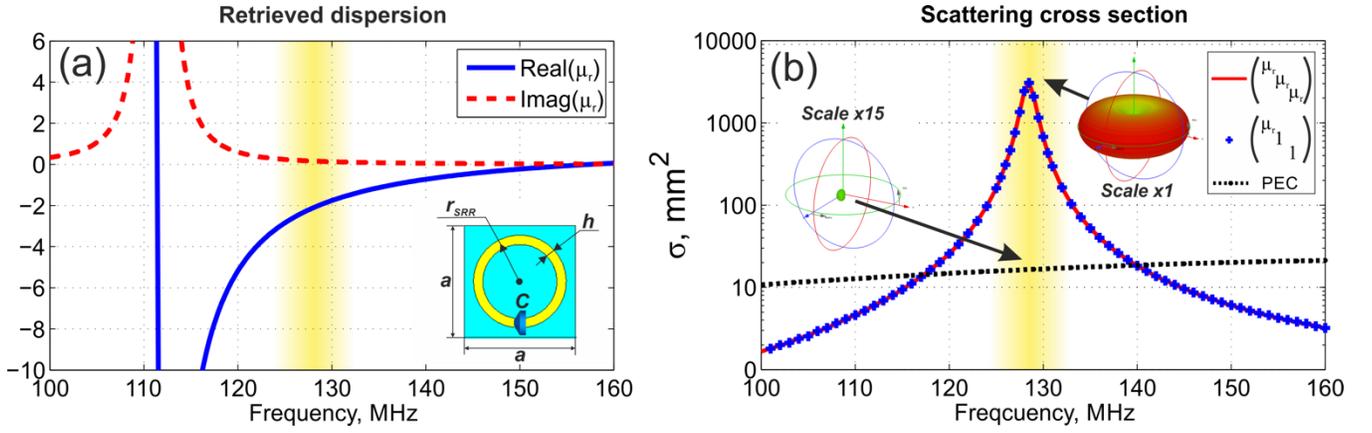

*Fig. 2.* (a) Dispersion of the homogenised array of split ring resonators (in the inset). Blue line – real part of the relative permeability, red dashed line – its imaginary part. Parameters of the array appear in the text. (b) Scattering cross section spectra of a 5 cm radius sphere made of a homogeneous material: red line – isotropic, blue dots – anisotropic, black dots – made of perfect electric conductor. Insets – scattering patterns of the structures (meta-sphere in 1:1 scale, PEC sphere in 1:15 scale for a better visibility on the linear scale).

**Experimental demonstration of artificial localized magnon resonances and super scattering at GHz spectral range**

In order to demonstrate the beforehand discussed phenomena, experimental studies will be performed at the GHz spectral range. The choice of the frequency band is solely pre-defined by the available experimental facilities (anechoic chamber). The same retrieval procedure was employed in order to extract and optimize effective parameters of SRRs arrays. In fact, exactly the same geometric parameters, as in the beforehand MHz case, were used apart from the lumped capacitor, which was taken here to be $C = 0.5$ pF. That is the reason why the same photograph appears on Fig. 1 and Fig. 4 for MHz and GHz respectively. Effective permeability dispersion appears in Fig. 3(a). In contrary to the MHz realization, deep subwavelength structuring at the GHz spectral range requires accurate miniaturizing of resonators and soldering of lumped elements, which also start influencing the scattering owing to their packaging sizes. As the result, the fabrication aspects set additional constraints on the dimensions of the subwavelength structure, which start experiencing retardation effects. The color map in Fig. 3(b) shows the evolution of the RCS spectrum as the function of the sphere size. The spectra were obtained numerically, while the particle effective tensor was taken to be anisotropic (hyperbolic dispersion). The white line, showing the main RCS peak, was calculated with Eq. 2. While this analytical formula was derived for the purely isotropic case, it fits extremely well the numerical data and shows the redshift of the peak with the radius increase. This effect will be clearly observed in the forthcoming experimental results.

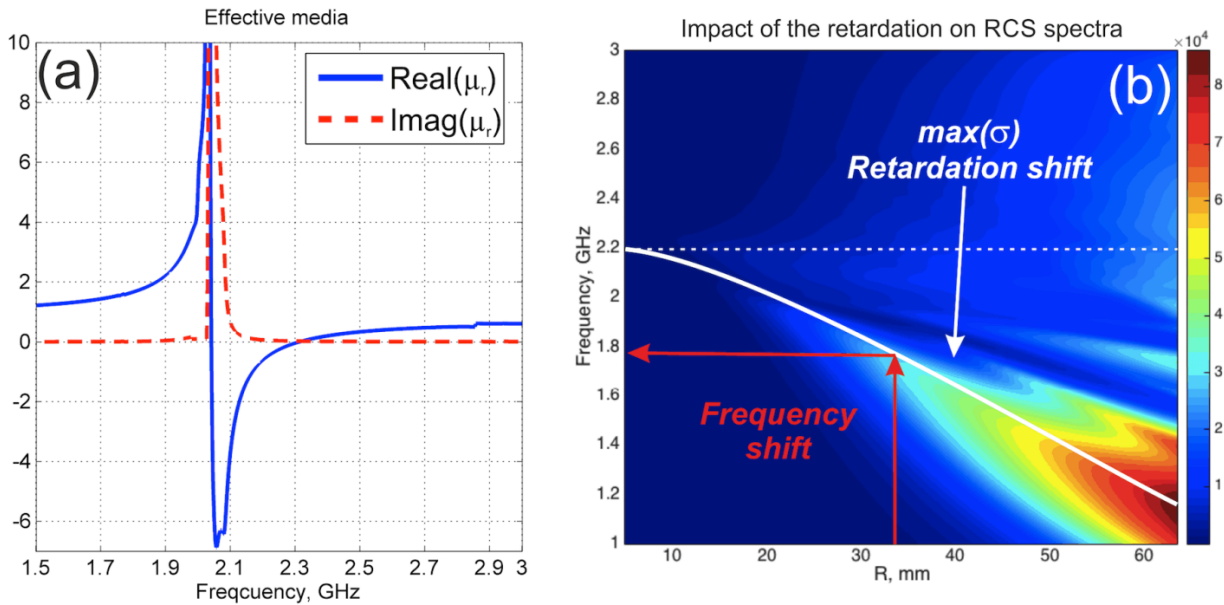

*Fig. 3 (a) Retrieved relative permeability spectrum. Red dashed line –imaginary part of $\mu_{eff}$, blue solid line - a real part of $\mu_{eff}$. (b) Color map of the RCS spectra (along the vertical axis) as the function of the sphere radius (horizontal axis). The sphere is made of a material with homogeneous anisotropic permeability with the dispersion of the nontrivial component, which appears on panel (a). White dashed line represents the position of the RCS peak without considering the retardation effect (Eq. 1). White solid line is calculated according to Eq. 2 and follows the RCS peak, underlining the influence of the retardation shift.*

Experimental sample, following the effective medium design, was fabricated to maintain the same geometric parameters. The meta-sphere contains 13 layers, separated by 5 mm distance (the period of the structure). SRRs arrays were fabricated on low loss *Isola* pc-board (parameters appear in the previous section*)* with a standard lithography, followed by chemical etching. Lumped elements (Multilayer Ceramic Capacitors 500R07S0R5AV4T) with a declared small tolerance (0.5%) in the nominal capacitance were soldered to the SRRs' gaps (542 elements in overall). The photograph of the sample appears in Fig. 1. Rectangular wideband horn antenna (IDPH2018) antenna connected to the transmitting port of a Vector Network Analyzer (VNA E8362B) was used as a plane wave excitation source. The scatterer was then located in the far-field of the antenna, at ~ 2.5 m distance, and a second identical horn antenna was used as a receiver to collect the scattered signal. Absolute RCS values were obtained with a set of standard calibrations, including the measurement a test metal sphere [37].

Fig. 4. shows the comparison between the prediction of the super scatterer's RCS and its actual measured performance. Panel (a) demonstrates numerically obtained RCS spectrum of the homogenized sphere with the anisotropic magnetic tensor. The full wave simulation of the entire structure, including all 13 planes with 542 SRRs with lumped capacitors, was performed and the resulting RCS appears in Fig. 4(b). The experimental data is presented in panel (c). It is worth noting the excellent agreement between all three methods. The main peak on the RCS spectrum corresponds to the dipolar resonance, which is calculated with Eq. 2 (recall the shift, originating from the retardation effect). The secondary peak, blue-shifted with respect to the main one, corresponds to the quadrupole mode. Its resonant condition is $Re(\mu_r) = -1.5$ in the fully quasistatic approximation. Its spectral location agrees with the dispersion curve of the effective medium (Fig. 3(a)). The maximum RCS value is as high as 250 cm$^2$, which is almost one order of magnitude larger than the geometric area of the scatterer ($\pi r^2$). This ratio can be further increased, if SRRs will be replaced with higher quality factor resonators, the substrate losses and nominal dispersion of lumped elements will be improved.

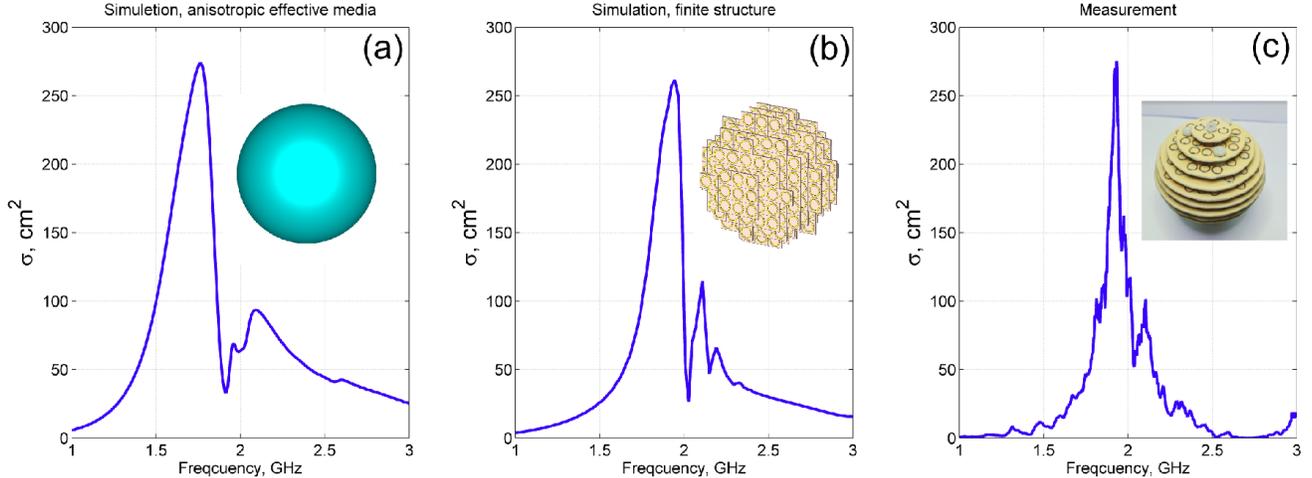

*Fig. 4. Scattering cross section spectra of metamaterial-based sphere with 33 mm radius. (a) Numerical modeling of the homogeneous anisotropic sphere. (b) Full wave numerical modeling of the metamaterial scatterer, made of split ring resonator arrays (13 planes, 542 SRRs in overall). (c) Experimental data for the sample, which follows the design of panel (b). Additional details appear in the main text.*

**Outlook and Conclusions**

The effect of artificial localized magnon resonance in arrays of near field coupled split rings was demonstrated. Negative effective permeability was shown to be responsible to the high scattering cross section of the metamaterial-based sphere. In particular, 5 cm radius (~1/50 fraction of a wavelength) scatterer can have a VHF radar signature, comparable with this of a big military aircraft. Experimental demonstration of the concept at the GHz frequency band was found to be in the extremely good agreement with theoretical predictions and full wave numerical simulations. Scattering cross sections of the designed objects, operating at the localized magnon resonance, were found to prevail the signatures of steel spheres and high quality factor ceramic resonators by orders of magnitude.

The demonstrated concept pays the way for further development of many applications, including short and long-range wireless communications, efficient remote RFID tags and their use in internet of things protocols, and many other areas.

**Acknowledgments**

This work was partly supported by PAZY foundation, BSF, and Kamin project.